# Vertical Profile Corrected Satellite NH$_3$ Retrievals Enable Accurate Agricultural Emission Characterization in China

Qiming Liu[1,2], Yilin Chen[3,*], Peng Xu[4], Huizhong Shen[1,2], Zelin Mai[1,2], Ruixin Zhang[1,2], Peng Guo[1,2], Zhiyu Zheng[1,2], Tiancheng Luan[1,2], Shu Tao[1,2,5,6]

[1]Shenzhen Key Laboratory of Precision Measurement and Early Warning Technology for Urban Environmental Health Risks, School of Environmental Science and Engineering, Southern University of Science and Technology, Shenzhen 518055, China

[2]Guangdong Provincial Observation and Research Station for Coastal Atmosphere and Climate of the Greater Bay Area, School of Environmental Science and Engineering, Southern University of Science and Technology, Shenzhen 518055, China

[3]School of Urban Planning and Design, Peking University, Shenzhen Graduate School, Shenzhen 518055, China

[4]Institute of Surface–Earth System Science, School of Earth System Science, Tianjin University, Tianjin 300072, China

[5]College of Urban and Environmental Sciences, Peking University, Beijing 100871, China

[6]Institute of Carbon Neutrality, Peking University, Beijing 100871, China

*Corresponding author, e-mail: ylchen2023@pku.edu.cn

## Abstract

Ammonia (NH$_3$) emissions significantly contribute to atmospheric pollution, yet discrepancies exist between bottom-up inventories and satellite-constrained top-down estimates, with the latter typically one-third higher. This study quantifies how assumptions about NH$_3$ vertical distribution in satellite retrievals contribute to this gap. By implementing spatially and temporally resolved vertical profiles from the Community Multiscale Air Quality model to replace steep gradients in Infrared Atmospheric Sounding Interferometer (IASI) retrievals, we reduced satellite-model column discrepancies from 71% to 18%. We subsequently constrained $NH_3$ emissions across China using a hybrid inversion framework combining iterative mass balance and four-dimensional variational methods. Our posterior emissions showed agreement with the a priori inventory (7.9% lower), suggesting that discrepancies between inventory approaches were amplified by overestimation of near-surface $NH_3$ in baseline satellite retrievals, potentially causing a 43% overestimation of growing season emissions. Evaluation against ground-based measurements confirmed improved model performance, with normalized root-mean-square error reductions of 1-27% across six months. These findings demonstrate that accurate representation of vertical profiles in satellite retrievals is critical for robust $NH_3$ emission estimates and can reconcile the long-standing discrepancy between bottom-up and top-down approaches. Our hybrid inversion methodology, leveraging profile-corrected satellite data, reveals that China's $NH_3$ emissions exhibit greater spatial concentration than previously recognized, reflecting agricultural intensification. This advancement enables timely and accurate characterization of rapidly changing agricultural emission patterns, critical for implementing effective nitrogen pollution control measures.

## 1. Introduction

Ammonia ($NH_3$) emissions play an important role in atmospheric pollution, primarily through their contribution to secondary fine particulate matter ($PM_{2.5}$) formation,[1] acid deposition, and ecosystem eutrophication.[2,3] These environmental impacts are associated with considerable public health concerns, as $PM_{2.5}$ derived from $NH_3$ has been linked to elevated risks of morbidity and mortality.[4] Prior studies suggest that reducing $NH_3$ emissions may offer a more cost-effective strategy for lowering $PM_{2.5}$ concentrations compared to controlling nitrogen oxides, with health benefits exceeding the costs of mitigation.[5]

China accounts for approximately one-sixth of global $NH_3$ emissions,[6] with agricultural activities—particularly livestock management and fertilizer application—being the dominant sources.[7, 8] These emissions exhibit notable seasonal and spatial variations, influenced by cropping cycles, meteorological conditions, and regional farming practices.[9-12] Despite their importance, $NH_3$ emissions remains difficult to quantify accurately. Bottom-up inventories continue to exhibit substantial uncertainties stemming from both activity data and emission factors,[8, 12, 13] with variations of annual emission estimates spanning up to 6 Tg.[6, 14] In particular, agricultural statistics often lack sufficient spatial and temporal resolution, and emission factors may not fully capture variability associated with climate, soil, and management practices.[13] As a result, annual emission estimates from different inventories can differ by more than 70%.[15]

To complement bottom-up approaches, observation-constrained inverse modeling has emerged as a promising alternative to improve emission estimates. While ground-based monitoring networks like AMoN-China have been utilized in inversions, they remain limited by sparse spatial coverage.[16] Satellite instruments—including the Infrared Atmospheric Sounding Interferometer (IASI),[17-20] Cross-track Infrared Sounder (CrIS),[21, 22] and Tropospheric Emission Spectrometer (TES)[15] — provide broader spatial coverage and have been increasingly used to estimate $NH_3$ emissions. However, comparisons between satellite-constrained and inventory-based estimates consistently reveal a systematic discrepancies, with top-down results typically 4%-63% higher than bottom-up values.[17, 21] These differences often exhibit a seasonal pattern, reaching up to 50% during summer months when agricultural emissions and volatilization intensify.[18, 21] While such gaps are frequently interpreted as evidence that bottom-up inventories may underestimate $NH_3$ emissions, they may also reflect unresolved uncertainties on the satellite retrieval side that have yet to be fully addressed.

Among the potential contributors to this discrepancy, the assumed vertical distribution of $NH_3$ in satellite retrievals is particularly important but often overlooked. The retrieval sensitivity of IASI increases with altitude due to favorable thermal contrast, making it highly dependent on vertical profile assumptions.[23, 24] Earlier retrieval schemes employed a Gaussian-shaped profile that declines by approximately 90% from the surface to 1.5 km altitude.[24-27] However, in-situ air craft and ground-based measurements suggest that such steep gradients may not be representative of actual conditions in many regions, particularly in areas with stronger vertical mixing or regional transport impacts.[28] This discrepancy may be especially pronounced in China, where intensive agricultural emissions, complex terrain, and diverse climate regimes result in heterogeneous vertical distributions that are not well captured by generic assumptions used in satellite retrievals. The recently released version 4 of the Artificial Neural Network for IASI (ANNI-v4) $NH_3$ retrieval incorporates total column averaging kernels (AVKs), which offer a means to evaluate and adjust for the effects of vertical sensitivity.[24] These AVKs make it possible to better account for the influence of assumed vertical profiles on retrieved columns and thereby reduce related uncertainties in top-down emission estimates.

In this study, we seek to address this issue by replacing the default vertical profiles used in IASI retrievals with spatially and temporally resolved profiles simulated by the Community Multiscale Air Quality (CMAQ) model. These profiles aim to better reflect regional variability in vertical $NH_3$ distributions, informed by emission sources, meteorological conditions, and transport processes. We then use the reprocessed IASI retrievals to constrain $NH_3$ emissions in China using a hybrid inversion framework that integrates the Iterative Mass Balance (IMB) approach and the four-dimensional variational (4D-Var) method.[19] This hybrid scheme combines the computational efficiency of IMB with the optimization accuracy of 4D-Var, allowing for high-resolution, long-term inversion applications. Our results indicate that incorporating more realistic vertical profiles can substantially reduce the discrepancy between satellite-based and bottom-up $NH_3$ estimates. The optimized emission estimates show more concentrated spatial patterns that may better reflect the increasingly clustered trend of agricultural practices in China. These improvements offer valuable insights for refining $NH_3$ emission inventories and advancing nitrogen pollution mitigation strategies.

## 2. Materials and Methods

### 2.1 Model Configuration and Simulation Framework

We employed the CMAQ v5.0.2 and its adjoint model for forward simulation and inverse modeling of $NH_3$ emissions.[29-31] Forward simulations were conducted to generate $NH_3$ column concentrations and vertical profiles for the entire year of 2017. However, calculating emission sensitivities during inventory optimization through inverse modeling requires substantial computational resources. To address this constraint, we strategically selected six representative months for simulation: January, April, July, and October to capture seasonal variability, as well as May and June to cover the peak agricultural period. Each simulation included a 10-day spin-up. The modeling domain over East Asia encompasses mainland China (Figure S1), with a horizontal resolution of 36 km by 36 km and 13 vertical layers extending up to approximately 20 km. CMAQ was configured with the CB05-AERO5 chemical mechanism and the ISORROPIA II for thermodynamic gas-particle partitioning of $NH_3$ and $NH_4^+$.[32, 33] The CMAQ adjoint model included multi-phase adjoints for gas-phase chemistry, aerosol formation, cloud processes, and atmospheric transport, enabling accurate sensitivity analysis of $NH_3$ total column concentrations with respect to emission perturbations. The adjoint system has previously demonstrated successful application in source attribution and emission optimization tasks.[19, 34]

Meteorological fields were generated using the Weather Research and Forecasting (WRF) Model v3.8.1, incorporating grid nudging with the Global Forecast System surface data from the National Centers for Environmental Prediction.[35] The WRF-simulated temperature, humidity, and wind fields were evaluated against observations from 412 surface meteorological stations (Figure S2).

### 2.2 IASI $NH_3$ Observations and Reprocessing

We utilized 2017 $NH_3$ observations from the IASI aboard the Metop-A satellite.[36]

Among the two daily overpasses (09:30 AM and 9:30 PM local solar time), only the 09:30 AM observations were used due to their lower uncertainty under favorable thermal contrast conditions.[19, 24] The analysis employed the IASI NH$_3$ Level 2 product version 4 (ANNI v4), which includes AVKs and associated vertical profile diagnostics that allow improved treatment of vertical sensitivity.[24]

Following Clarisse et al.,[24] we reprocessed the NH$_3$ column densities by replacing the a priori profile used in the retrieval with CMAQ-simulated vertical profiles. This procedure enables a more direct comparison between satellite retrievals and model-simulated column densities. The reprocessed column $X^m$ was calculated following equation (1) - (3) as:

$$X^m = \frac{X^a}{\sum_z A_z^a m_z} \quad (1)$$

$$A_z^a = \frac{1}{N} \frac{X^a}{X^{|z}} \quad (2)$$

$$m_z = \frac{M_z^m}{M^m} \quad (3)$$

where $X^a$, $X^{|z}$, and $N$ are provided in the IASI-ANNI-v4 product, representing the retrieved column based on the original a priori profile, the retrieved total column assuming all NH$_3$ concentrated at level $z$, and the normalization factor, respectively. $A_z^a$ thus represents the normalized AVK, reflecting the sensitivity of the retrieved column to variations at each altitude. The term $m_z$ denotes the fractional distribution of the CMAQ-simulated NH$_3$ column in vertical layer $z$, calculated as the ratio of the partial column in that layer to the total simulated column. Because the vertical layers of IASI and CMAQ differ, we applied a mapping between the IASI pressure levels and CMAQ model layers before reprocessing the retrievals. This mapping was performed by interpolating CMAQ profiles to the IASI pressure grid using pressure midpoints.

Post-filtering was applied based on the reprocessed $X^m$ following the criteria recommend by Clarisse et al.[24] (Text S1) The reprocessed data were then mapped onto the CMAQ grid by calculating the hourly arithmetic mean of all observations within each grid cell. For uncertainty quantification, we derived the total uncertainty estimates corresponding to the averaged column densities in each grid cell by combining reported random and systematic errors, without including the vertical profile uncertainty, following equation (4),

$$\sigma_{\bar{x}}^2 = \frac{1}{n^2}\sum_{i=1}^{n}\sigma_{rX_i}^2 + \frac{1}{n}\sum_{i=1}^{n}\sigma_{sX_i}^2 \quad (4)$$

where $\sigma_{\bar{x}}$ is the average error at a grid point, n is the number of measurements within that grid, and $\sigma_{rx,i}$ and $\sigma_{sx,i}$ are the random and systematic errors for the i-th measurement, respectively. The total uncertainties are used to construct the error covariance matrices for IASI NH$_3$ retrievals.

**2.3 Prior NH$_3$ Emission Inventory**

The prior NH$_3$ emission inventory integrates multiple datasets to characterize seasonal and spatial variability across China. Monthly agricultural emissions from fertilization application and livestock management were derived from high-resolution, sector-specific inventories based on city or county-level activity data and environmental condition-adjusted emission factors.[11, 37] These emission factors consider parameters such as soil acidity, meteorological conditions, and agriculture practices. The aggregated emission estimates were further allocated at 1 km by 1 km resolution using land use and rural population as spatial surrogates. In this study, we aggregated these gridded emissions to the CMAQ 36 km by 36 km grids. Temporal interpolation was performed using the AiMa emission inventory framework. Non-agricultural NH$_3$ emissions and emissions for other species from transportation, power generation, residential, and industrial sectors were also obtained from the AiMa inventory,[38] which has been applied in air quality modeling and forecasting studies in China.[39] Emissions for northern India were derived from the Intercontinental Chemical Transport Experiment-Phase B emission inventory.[40]

**2.4 Hybrid Inversion System**

To optimize $NH_3$ emission estimates, we applied the hybrid inversion framework developed by Chen et al.,[19] which combines the IMB method with a 4D-Var assimilation scheme. This two-step approach improves computational efficiency while maintaining the spatial resolution of the final posterior inventory.

In the first step, IMB adjusts emissions by iteratively scaling them according to the ratio of observed to simulated monthly average $NH_3$ columns in coarse 216 km by 216 km moving windows until convergence is achieved (Text S2). This intermediate product is used as the initial emission estimates for the 4D-Var inversion performed at daily temporal and 36 km by 36 km spatial resolution. The 4D-Var method minimizes a cost function $J$ defined as:

$$J = \gamma(\varepsilon - \varepsilon_o)^T S_a^{-1}(\varepsilon - \varepsilon_o) + (\Omega_o - F(\varepsilon))^T S_o^{-1}(\Omega_o - F(\varepsilon)) \quad (5)$$

where $\varepsilon$ is the emission scaling factor vector, $S_a$ and $S_o$ are the error covariance matrices for the prior and observations, respectively. Error covariance matrices were assumed to be diagonal with prior emission uncertainty set to 100% and observation uncertainty derived from IASI error estimates.[41] $F$ is the forward model, and $\Omega_0$ denotes the reprocessed IASI observations. A regularization parameter $\gamma$ is selected based on the L-curve method (Figure S3).[42] The cost function was minimized using the L-BFGS-B algorithm, with gradients calculated by the CMAQ adjoint model. IMB convergence was defined as a <10% change in normalized root mean square error (NRMSE), and 4D-Var convergence was defined as a <2% change in J or attainment of a local minimum.

**2.5 Evaluation Methods**

**2.5.1 Satellite Evaluation**

Model performance was evaluated by comparing CMAQ-simulated $NH_3$ columns with the reprocessed IASI retrievals. Statistical metrics including normalized mean bias (NMB), NRMSE, and Pearson correlation coefficient (r) were calculated to quantify agreement and improvements post-optimization.

**2.5.2 Surface Observation Evaluation**

To further assess the posterior emissions, simulated surface $NH_3$ concentrations were compared with observations from the AMoN-China,[43] which includes 53 monitoring stations across the country (Figure S1) using standardized passive sampling techniques. Due to data availability, we compared 2017 model outputs with measurements collected between September 2015 and August 2016, recognizing this temporal mismatch as a potential limitation.

**2.5.3 Spatial Inequality Analysis**

To evaluate spatial concentration patterns of $NH_3$ emissions, we computed Theil's T index, a measure of distributional inequality:

$$T_T = \frac{1}{n}\sum_{i=1}^{n}(\frac{E_i}{E_{mean}} \times \log(\frac{E_i}{E_{mean}})) \quad (6)$$

Where $n$ denotes the total number of grid cells in China, $E_i$ signifies the monthly emission at grid cell $i$, and $E_{mean}$ corresponds to the monthly mean emission across all grid cells. The $T_T$ index quantifies the spatial heterogeneity of emission distribution patterns, where $T_T$ equals zero when emissions exhibit an even distribution and higher values indicates a more concentrated spatial distribution.

## 3. RESULTS AND DISCUSSION

**3.1. Comparing $NH_3$ Emission Estimates from Top-Down and Bottom-Up Inventories**

The a priori $NH_3$ emission inventory used in this study estimates total emissions in

China at 12.1 Tg for 2017. Emissions are primarily attributed to livestock management (59.3%) and fertilizer application (34.2%), with smaller contributions from industrial processes (2.5%), residential activities (2.6%), and ground transportation (0.4%). When compared to other published inventories (Figure 1a),[6, 8, 14, 16, 17, 44-51] the a priori inventory used in this study shows broadly consistent source sector contributions and aligns well with recently updated bottom-up estimates such as CEDS and HTAP_v3,[48, 50] However, the total emissions in this inventory are approximately 9% higher than the multi-inventory average for 2017 (11.1 Tg). This difference is largely attributable to agricultural sources, for which the a priori inventory estimates are 28-69% higher than those reported in widely used inventories such as MEIC v1.4,[47] EDGAR,[6] PKU-$NH_3$.[49]

A more pronounced discrepancy emerges when comparing our a priori inventory with satellite-constrained top-down estimates. Our 2017 total is 27% lower than the average of top-down estimates (16.7 Tg). Statistical comparisons using independent two-sample t-tests ($p < 0.001$, N = 31) across the 2016-2018 period confirm a consistent and significant underestimation by bottom-up approaches relative to top-down methods. Similar discrepancies have been documented in earlier studies, indicating enduring methodological gaps in reconciling these two approaches.[15-18, 21] Seasonally, the divergence between bottom-up and top-down estimates displays a distinct temporal pattern (Figure 1b).[8, 17, 18, 47-49, 51] The largest relative discrepancy occurs in summer, when bottom-up estimates (4.0 Tg) are 29% lower than top-down estimates (5.6 Tg), followed by winter and spring, with differences of 27% and 24%, respectively. In contrast, fall shows near agreement, with only a 2% difference. July exhibits the largest absolute difference (0.9Tg), coinciding with peak agricultural activities and highest ambient temperatures across much of China.[8, 15, 37]

The persistence of this gap, even in the context of recent inventory improvements, suggests that unresolved uncertainties beyond direct emission estimates may be contributing to the divergence. These include, notably, the treatment of vertical $NH_3$ distribution in satellite retrievals and limitations in the retrieval algorithms themselves.

While prior efforts have focused on refining retrieval accuracy,[25-27] the introduce of total column AVKs in the latest IASI NH$_3$ product (ANNI v4) now enables post-retrieval adjustment using externally derived vertical profiles.[24] In this study, we use CMAQ-simulated NH$_3$ vertical profiles to reprocess IASI retrievals and assess the role of vertical distribution assumptions in shaping emission estimates. Addressing this factor is essential for reducing uncertainty in satellite-constrained inversions. The persistent gap between bottom-up and top-down NH$_3$ estimates carries important implications for nitrogen deposition budgets and air quality management in China, particularly in regions with intensive agriculture and high population exposure.[52-55]

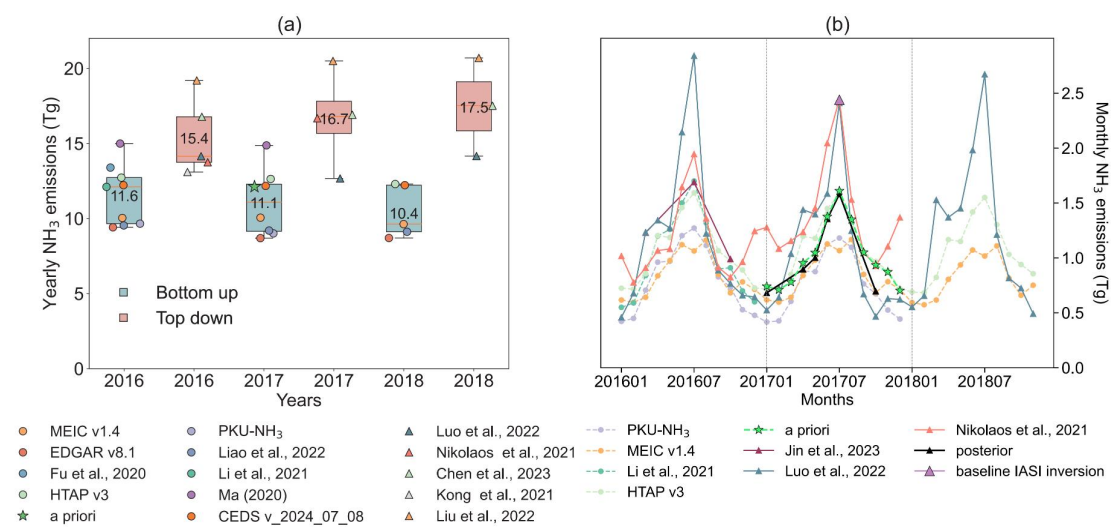

**Figure 1. Comparison of NH$_3$ emission estimates from bottom-up and top-down inventories across China (2016-2018). (a)** Boxplots of annual total NH$_3$ emissions from multiple bottom-up (blue) and top-down (red) inventories. Values represent multi-year averages for each inventory. **(b)** Monthly NH$_3$ emission trends showing seasonal cycles throughout the study period. Various markers represent different inventory sources as indicated in the legend.

## 3.2. Impact of Replacing the CMAQ NH$_3$ Vertical Profile

Assumptions regarding the vertical distribution of NH$_3$ in satellite retrievals represent a significant source of uncertainty contributing to the discrepancy between bottom-up and top-down NH$_3$ emission estimates. Analysis of IASI AVKs confirms that retrieval sensitivity to NH$_3$ increases with altitude, particularly in the mid- to upper troposphere, where thermal contrast conditions are more favorable (Figure S4b).[24] This altitude-dependent sensitivity, combined with simplified assumptions about vertical distribution, can lead to systematic biases in emission estimates derived from satellite observations. While prior $NH_3$ generally assume high concentrations near the surface with rapid vertical decay,[24, 56, 57] these profiles may not adequately capture actual atmospheric $NH_3$ distributions in regions with strong vertical mixing or long-range transport. The IASI retrieval algorithm, for instance, uses a baseline vertical profile that declines by approximately 90% from the surface to 1.5 km—the typical planetary boundary layer (PBL) height (Figure 2a).

Comparative analysis with in-situ observations in high-emission regions of China (Beijing[58] and Baoding[59]) reveals that the IASI baseline profile substantially overestimates the vertical concentration gradient (Table S1). At the Baoding site, the 0.5 km-to-1.5 km $NH_3$ concentration ratio is approximately 4.8:1 in the IASI a priori profile, compared to a more modest ratio of approximately 1.4:1 from aircraft measurement. Although CMAQ-simulated profiles also exhibit steeper gradients than observed, their bias is less severe, with a ratio of 2.4:1 at the same location. Over mainland China, the CMAQ profiles predict 27%-90% higher $NH_3$ concentrations above 1.5 km than the IASI baseline profile. These differences are critical, as they directly affect satellite retrievals. However, such comparisons remain limited by the scarcity of vertical in situ measurements, underscoring the need for more comprehensive observational datasets to improve model representation and satellite retrieval accuracy.

To reduce the influence of unrealistic profile assumptions, we reprocessed the IASI $NH_3$ retrievals by replacing the baseline vertical profile with spatially and temporally varying profiles from CMAQ (Section 2.2). These model-derived profiles reflect localized emission patterns, meteorological dynamics, and vertical transport processes.[60-62] This adjustment leads to a notable reduction in the discrepancy between satellite-derived $NH_3$ columns and those simulated based on the a priori emission inventory. On average, original IASI retrievals using the baseline profile are 71% higher than bottom-up estimates, with the largest differences observed during the spring and summer (Figure 2b). During these seasons, CMAQ simulates enhanced NH$_3$ concentrations above the PBL (Figure S4a), which can be attributed to active vertical mixing and convective uplift, as reported in previous studies.[59, 62-64] After incorporating the CMAQ-derived vertical profiles, the annual-averaged difference between satellite and model columns decreases to 18%. Although some underestimation remains during the growing season (April to July), the bias magnitude is reduced by 81%.

Regionally, the reprocessing of IASI NH$_3$ retrievals reduces the gap between IASI-retrieved and bottom-up NH$_3$ columns across six major Chinese regions (Figure 2c-h). The largest improvements are observed in Northwest and Southwest China, where local emissions are relatively low but NH$_3$ concentrations above the PBL remain substantial. In these regions, CMAQ simulations show that more than 40% of the column NH$_3$ resides above the PBL—far higher than the 5% assumed in the baseline IASI profile. This discrepancy explains the overestimation of emissions in these regions reported in earlier top-down studies.[18] With profile replacement, the relative difference between retrieved and bottom-up columns decreases from 298% to 123% in Northwest China and from 185% to 30% in Southwest China. In contrast, Northeast China shows smaller improvements, as stronger surface emissions result in vertical profiles that already resemble the baseline assumption, with most NH$_3$ confined within the PBL.

To further examine the role of regional transport in shaping elevated NH$_3$ profiles, we conducted a sensitivity simulation by excluding emissions from India and domain boundary inflows (Text S3). The resulting 24-percentage-point decrease in NH$_3$ above the PBL in Northwest and Southwest China confirms that regional transport enhances elevated NH$_3$ levels (Figure S5). This is consistent with airborne measurements from Pu et al.,[59] which documented NH$_3$ concentrations up to 20 ppb at 2500-3000 m during periods of strong regional transport.

Overall, the comparison of gridded annual-averaged NH$_3$ columns shows improved consistency between satellite-derived and simulated NH₃ fields following profile correction. The NMB improves from -45% to -4.7%, and the NRMSE decreases from 71% to 66%. These results demonstrate that vertical profile assumptions constitute a key source of uncertainty in satellite-based NH₃ emission estimates, particularly in regions and seasons where vertical transport leads to elevated NH₃ layers above the boundary layer.

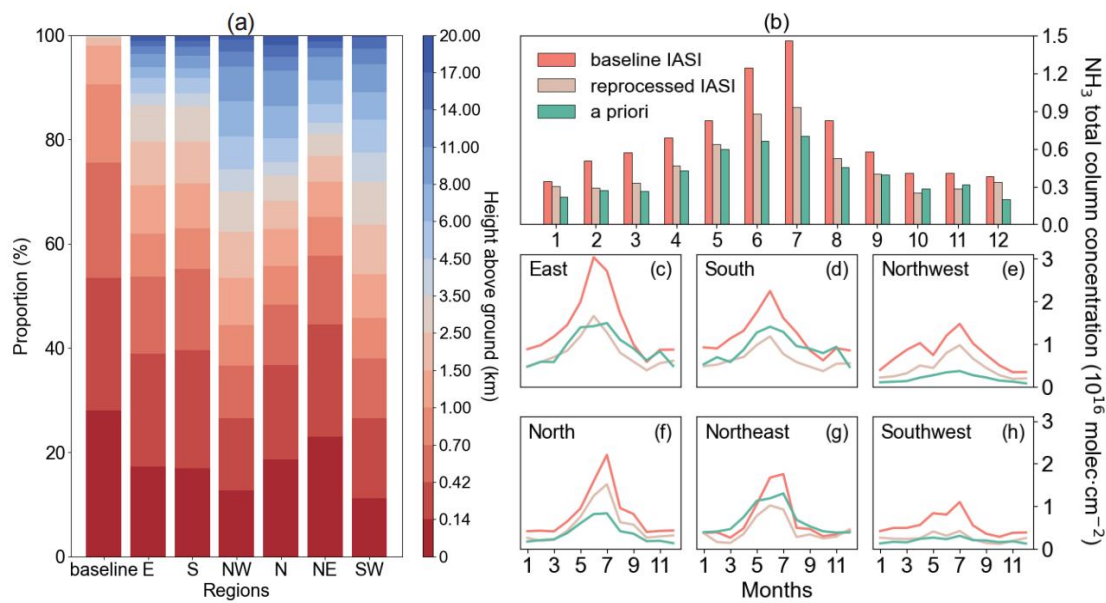

**Figure 2. Vertical NH₃ profile distributions and their impact on total column estimates.** **(a)** Vertical NH₃ distribution showing the average fractional contribution (%) of each altitude layer to the total column across six Chinese regions (E: East, S: South, NW: Northwest, N: North, NE: Northeast, SW: Southwest), along with the baseline IASI vertical profile. (b) Monthly mean NH₃ total column concentrations over China from baseline IASI retrievals, reprocessed IASI using CMAQ profiles, and model simulations using the a priori inventory. **(c–h)** Monthly mean NH₃ total column concentrations in the six regions. Regional boundaries follow the classification from the Resource and Environmental Science Data Center (Figure S1).[65]

### 3.3. Evaluation of NH₃ Emission

We used the reprocessed IASI NH₃ observations as constraints in our inversion framework to derive optimized NH₃ emission estimates for China. The posterior emission estimates preserve the seasonal cycle represented in the a priori inventory but exhibit notable regional and monthly adjustments. Estimated emissions for January, April, July, and October were 0.68 Tg, 0.89 Tg, 1.57 Tg, and 0.70 Tg, respectively. Additional estimates for May (1.00 Tg) and June (1.36 Tg) were included to capture the full progression of the growing season, when agricultural $NH_3$ emissions typically peak.

On average, posterior emissions estimates are 7.9% lower than the a priori values, with monthly differences ranging from -24% to 1.8%. The overall seasonal trend remains consistent, with the ratio between peak emissions in July and the January minimum reaching 2.3 in the posterior inventory, closely matching the ratio of 2.2 in the a priori inventory. In comparison, this ratio varies from 1.7 to 4.6 across other bottom-up and top-down inventories.[8, 17, 47-49, 51] Differences in peak-season emissions remain a key source of variability among inventories. Our posterior estimate for the growing season totals 4.82 Tg, closely matching the a priori estimate of 4.98 Tg in both magnitude and monthly distribution. This agreement stands in contrast to prior inversion studies, which reported pronounced low biases in bottom-up inventories during the growing season.[17, 18, 51] Notably, our July posterior estimate differs from the a prior value by only 1.8%, whereas previous studies have reported discrepancies as large as 66%.[17, 51]

Further analysis indicates that much of the historical discrepancy between bottom-up and top-down $NH_3$ estimates may stem from overestimation of near-surface $NH_3$ in the IASI baseline vertical profile. In our July 2017 test inversion using baseline IASI retrievals, the estimated emission reaches 2.4 Tg, closely matching previous top-down values (Figure 1b).[17, 51] Although our analysis is limited to July, the high bias observed in the baseline retrievals is likely to persist across other warm-season months when increased surface heating and stronger vertical gradients promote the upward transport of $NH_3$ into layers where satellite sensitivity is enhanced. If this bias extends throughout the growing season, our results suggest that prior top-down estimates may have overestimated emissions by as much as 43% (2.1 Tg) relative to our profile-corrected posterior inventory.

Spatially, the posterior emission inventory identifies the same major hotspots as the a priori inventory, including the North China Plain (NCP), the Sichuan Basin, and the edge of the Ili River Valley in Xinjiang (Figure 3a). However, differences in emission magnitudes reveal a more concentrated pattern in these regions, alongside reduced emissions across southern, southwestern, and eastern China. For instance, $NH_3$ emissions over the NCP during the growing season are 10% higher in the posterior estimate, contributing over 25% of national emissions. In the Ili River Valley, emissions are 162% higher than in the a priori inventory, likely reflecting increased contributions from livestock management and oasis agriculture in recent years.[8, 9] This trend is consistent with the high $NH_3$ emissions over the NCP and the northwestern China reported by Chen et al.[20] In addition to these inter-regional contrasts, the posterior inventory reveals sharper intra-provincial disparities between high- and low-emission zones. In Sichuan and Hebei provinces, for example, the a priori inventory underestimates emissions in the core of the Sichuan Basin and northern Hebei — areas with intensified agricultural production—while overestimating emissions in surrounding regions (Figure 3b). Quantitatively, emissions in the central Sichuan Basin are 81% higher in the posterior inventory, while surrounding areas exhibit 56% lower emissions. These findings underscore the significant spatial heterogeneity in agricultural emissions that may be obscured in conventional bottom-up inventories.

While the spatial locations of emission hotspots remain broadly consistent, the intensity contrasts between major agricultural centers and other regions, as well as within provinces, differ considerably. These patterns suggest that the spatial allocation methods used in the a priori inventory may not fully capture the increasingly centralized nature of agricultural activities in China. Recent studies have reported rapid growth in large-scale farming operations and confined animal feeding operations,[66-68] trends that are difficult to reflect using conventional inventory approaches reliant on coarse administrative statistics. Between 2014 and 2022, the number of large-scale livestock facilities increased by 39%,[68] indicating a shift from dispersed smallholder production to concentrated industrial-scale systems. However, regular updates to high-resolution inventories remain constrained by limited access to detailed activity data and farm-specific emission factors. Our results highlight the potential of hybrid inversion approaches, supported by high-resolution satellite data, to bridge these gaps and more accurately represent evolving emission patterns.

To evaluate the spatial inequality of emissions, we examined the cumulative distribution of $NH_3$ emissions and grid cell area, sorted by emission intensity (Figure 3c). Throughout the growing season, the top 10% of high-emitting grids account for nearly half of the national $NH_3$ emissions. Compared to the a priori inventory, the posterior inventory shows a greater degree of spatial concentration in April, May, and June. The contribution of the top 10% grids increases from 46%, 47%, and 47% to 56%, 56%, and 54%, respectively. We quantified this trend using Theil's T index (Section 2.5.3),[69] which confirmed a substantial increase in spatial concentration, with index values rising from 0.83 to 1.04 in April, 0.84 to 1.06 in May, and 0.85 to 1.02 in June. In contrast, July shows little change, with the Theil's T index shifting slightly from 0.88 to 0.87, indicating already high levels of spatial inequality during peak emissions. Given the short atmospheric lifetime of $NH_3$, its environmental impacts on $PM_{2.5}$ formation and nitrogen deposition are primarily local.[70] The more concentrated emission pattern identified in the posterior inventory implies greater environmental burdens in hotspot-adjacent regions than previously indicated by the a priori inventory. This localized effect underscores the importance of targeting emission control strategies in high-emitting regions, where interventions would yield disproportionate benefits for improving regional air quality and reducing nitrogen deposition.[71]

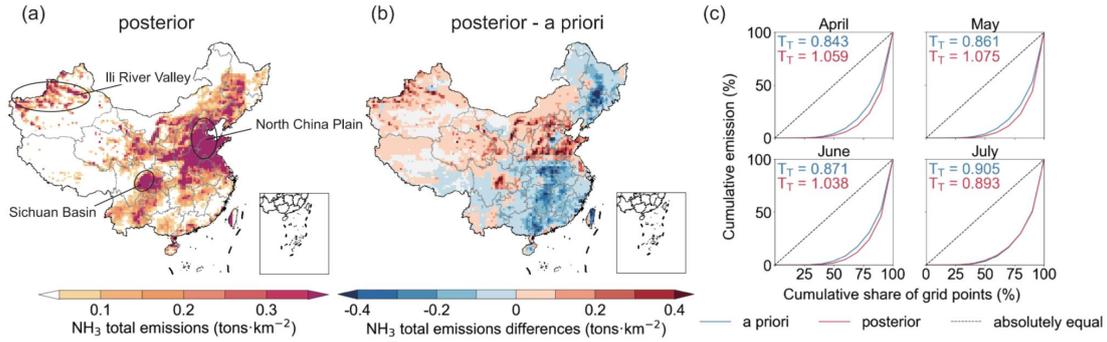

**Figure 3. Spatial patterns of NH$_3$ emissions during the growing season (April – July). (a)** Posterior NH$_3$ emission intensity distribution across China, with key agricultural regions labeled. **(b)** Spatial differences between posterior and a priori emission inventories. **(c)** Cumulative emission distribution curves showing the relationship between cumulative percentage of grid cells (sorted by increasing emission intensity) and cumulative percentage of total emissions. Theil's T indices (T$_T$) quantify spatial concentration for both a priori (blue) and posterior (red) inventories.

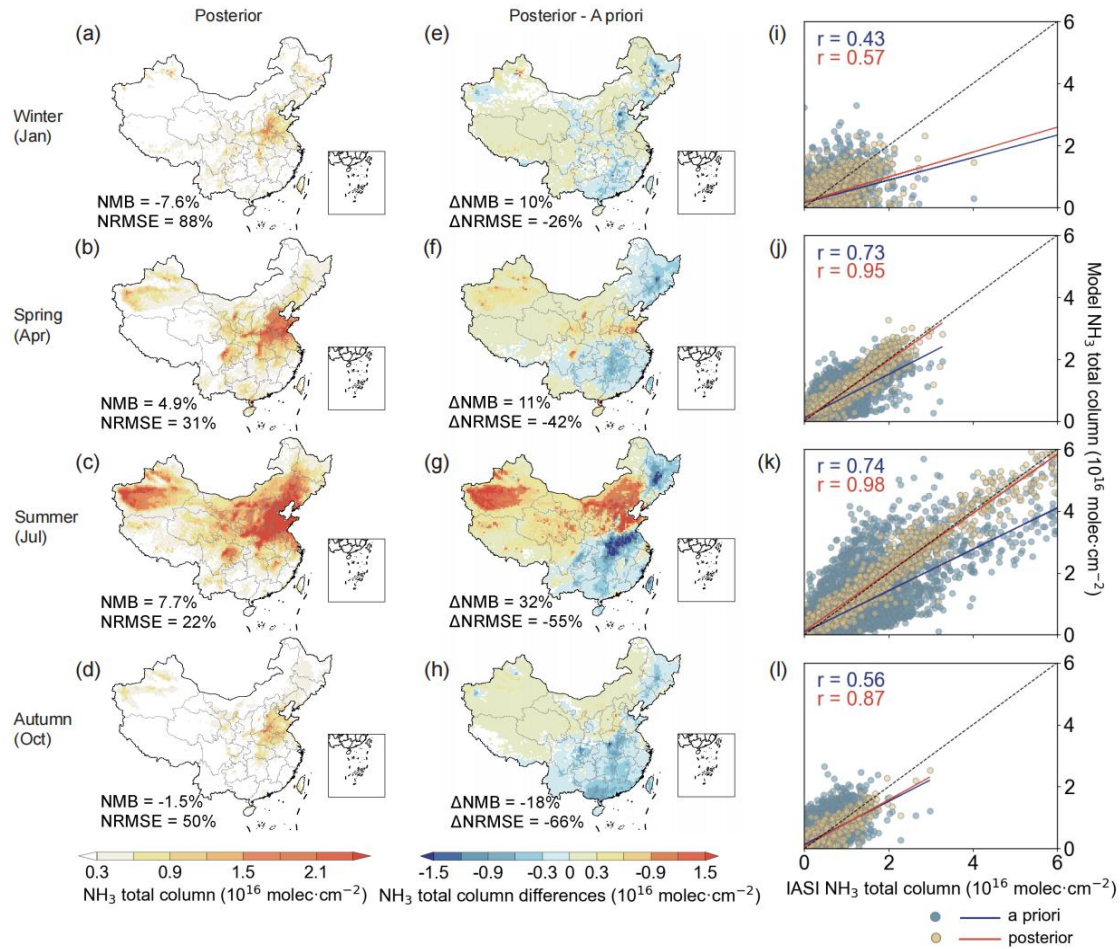

**Figure 4. Seasonal NH$_3$ total column density patterns from CMAQ simulations and model evaluation. (a–d)** CMAQ-simulated NH$_3$ column densities using posterior emission inventory across China for representative months of each season, with normalized mean bias (NMB) and normalized root-mean-square error (NRMSE) relative to reprocessed IASI observations shown. **(e–h)** Spatial differences between CMAQ simulations using posterior and a prior emission inventories, with corresponding changes in performance metrics (ΔNMB and ΔNRMSE). **(i–l)** Scatter plots comparing reprocessed IASI NH$_3$ columns with CMAQ simulations using a priori (blue) and posterior (red) emission inventories, with Pearson correlation coefficients (r) for each simulation.

### 3.4. Hybrid Retrieval Results and Ground Validation

Evaluation of CMAQ-simulated NH$_3$ column densities against reprocessed IASI NH$_3$ observations indicates that the inversion significantly improved model-satellite agreement across China, particularly during the growing season (Figure 4 and S6). Prior to optimization, simulations exhibited substantial spatial discrepancies relative to the reprocessed satellite data, with a NRMSE of 66%. Following optimization, model biases were substantially reduced and spatial correlations improved in all seasons. The NMB approached zero, confirming the technical success of the inversion in adjusting emissions to align with satellite constraints.

The adjustments were spatially heterogeneous. As shown in Figure 4 (middle column), simulated column densities based on posterior emissions increased emissions in northwestern and northern China—especially in Xinjiang—while decreasing across large areas of southern and northeastern China. In regions with high-emission intensities (defined as monthly emission rates > 0.3 tons/km$^2$), the inversion notably improved model performance. For July, the NMB decreased from 11% to 2.8%, the NRMSE declined from 57% to 14%, and the r increased from 0.74 to 0.98. Optimization performance varied seasonally, with the most substantial improvements observed in July (70% reduction in NRMSE) and the least improvement in January (22% NRMSE reduction). This seasonal dependence is consistent with known limitations of satellite constraints during winter, when weaker thermal contrast and

lower ambient $NH_3$ concentrations results in higher relative undertainties.[24, 72]

The optimized inventory was further evaluated against independent surface measurements from the AMoN-China network, which includes 53 monitoring sites across the country.[43] Across all evaluated months, the posterior simulations demonstrated improved agreement with ground-based observations, reducing monthly NRMSE by 1%-27% and increasing correlation coefficients in all months except January (Figure 5 and S7). July showed the most substantial improvement, particularly in high-emission regions, where the NRMSE decreased by 32% compared to a 27% domain-wide average, accompanied by consistent improvements in NMB and r values (Figure S8).

These results confirm that the inversion effectively refined the spatial distribution of emissions in critical hotspot areas. For example, at the LCA site in the North China Plain, the difference between observed and simulated surface $NH_3$ concentrations were reduced by 63% (Figure 5c,g), following a downward adjustment in posterior emissions in that grid cell. This correction aligns with reports that livestock facilities in this region adopt advanced management practices that reduce $NH_3$ emissions compared to conventional inventory assumptions.[73] The inversion effectively identified and corrected for such overestimates, demonstrating the ability of the hybrid framework to resolve finer-scale spatial emission structures in densely populated agricultural zones. Despite the overall improvement, systematic underestimation persisted in areas with lower emission (defined as monthly emission rates < 0.05 tons/km$^2$) rates. This limitation is partly attributable to greater relative uncertainties in both satellite retrievals and ground-based measurements under low $NH_3$ conditions, which limits the capacity of the inversion to constrain emissions in these regions.[43]

The evaluation results highlight the importance of vertical profile correction in improving top-down emission estimates. A parallel inversion using baseline IASI $NH_3$ retrievals produced poorer agreement with surface observations in July, with a NRMSE of 88% compared to 76% for the profile-corrected posterior inventory

(Figure S9). At high-emission sites, the NMB decreased from 42% with the baseline retrieval inversion to -3% in the optimized inventory. This substantial reduction in positive bias can be attributed to correcting the artificially elevated emission estimates caused by the steeper vertical gradients assumed in the baseline retrievals. These findings confirm that integrating CMAQ-modeled $NH_3$ vertical profiles into satellite retrievals substantially improves the accuracy of surface-level $NH_3$ representation.

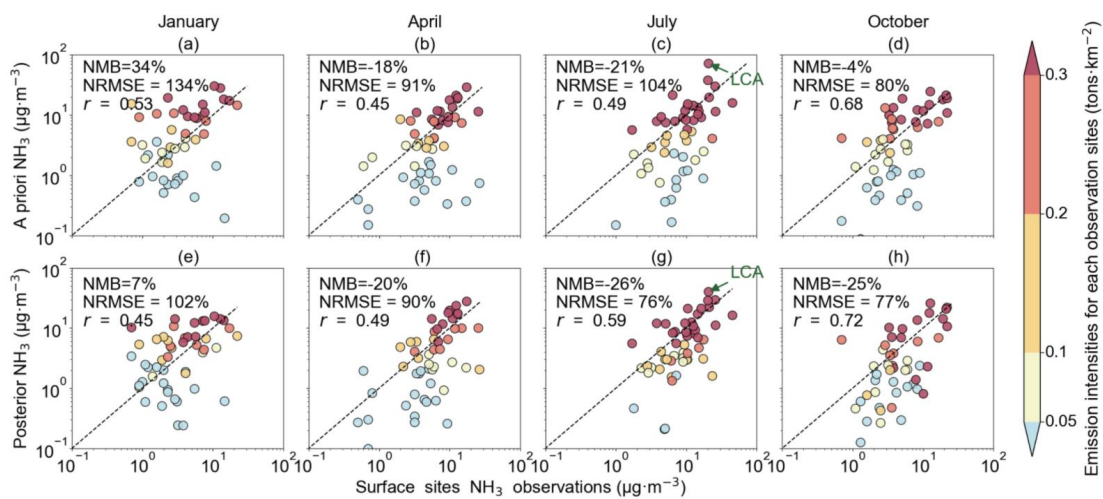

**Figure 5.** Evaluation of CMAQ simulated surface $NH_3$ concentrations against ground observations. **(a–d)** Scatter plots comparing $NH_3$ concentrations simulated using a priori emissions versus AMoN-China surface observations for January, April, July, and October. **(e–h)** Corresponding comparisons using posterior emission inventory. Point colors indicate local emission intensity at each observation site. Performance metrics (normalized mean bias, NMB; normalized root-mean-square error, NRMSE; and Pearson correlation coefficient, r) are shown in each panel. "LCA" indicates the Luancheng Agricultural station.

**Implications**

This study addresses a fundamental challenge in atmospheric $NH_3$ quantification by reconciling the persistent gap between bottom-up and top-down emission estimates through eliminating the impact of vertical profile assumption. By replacing simplified baseline profiles with spatially and temporally resolved CMAQ-simulated vertical distributions, we demonstrate that vertical profile assumptions significantly contribute to discrepancies in satellite-derived emission estimates. Our approach reduced the difference between satellite and model columns from 76% to 21% for the growing season, suggesting that prior top-down studies may have overestimated peaking emissions by up to 43 % due to profile-related biases. The findings reveal increasingly concentrated spatial emission patterns that better reflect China's agricultural intensification trends, with the top 10% of high-emitting grids accounting for over half of national $NH_3$ emissions. These improved emission characterizations have important implications for targeted pollution control strategies, as they indicate greater environmental burdens in hotspot-adjacent regions than previously recognized. The methodology presented here offers a valuable framework for enhancing satellite-based emission inventories that can be extended to other regions facing similar challenges in agricultural emission quantification.

**Limitations**

The attribution analysis in this study advances vertical profile representation in satellite retrievals but faces challenges due to limited vertical $NH_3$ measurements. While we show improvements over baseline retrievals in the two sampling sites, characterizing vertical transport across diverse landscapes in China remains challenging. More extensive aircraft measurements and vertical profiling campaigns are needed to further refine these representations, particularly in regions with complex atmospheric circulation patterns. Satellite retrieval performance varies seasonally, with winter estimates hampered by weaker thermal contrast conditions. The persistent challenges in low-concentration environments suggest that future satellite instruments with enhanced signal-to-noise ratios could substantially improve emission quantification in transitional regions between agricultural hotspots and background areas.

**Data availability**

Input datasets related to this paper are publicly available. The IASI/Metop-A $NH_3$ total column Level 2 product (IASI v4.0.0) is available from the AERIS IASI portal:

https://iasi.aeris-data.fr/NH3_IASI_A_data. Meteorological fields were generated using WRF v3.8.1 with grid nudging based on Global Forecast System (GFS) surface data from the National Centers for Environmental Prediction (NCEP) available at https://www.nco.ncep.noaa.gov/pmb/products/gfs/#GFS. Surface $NH_3$ observations were obtained from the AMoN-China network, with data access procedures as described by Pan et al.[43]

**Code availability**

The CMAQ Adjoint 5.0 model code an be accessed at https://github.com/USEPA/CMAQ_ADJOINT (https://doi.org/10.5281/zenodo.3780216). MATLAB R2021a was used for source attribution analysis in this study.

# References

(1) Hodan, W. M.; Barnard, W. R. Evaluating the Contribution of $PM_{2.5}$ Precursor Gases and Re-entrained Road Emissions to Mobile Source $PM_{2.5}$ Particulate Matter Emissions. 2004.

(2) Guo, J. H.; Liu, X. J.; Zhang, Y.; Shen, J. L.; Han, W. X.; Zhang, W. F.; Christie, P.; Goulding, K. W. T.; Vitousek, P. M.; Zhang, F. S. Significant Acidification in Major Chinese Croplands. *Science* **2010**, *327* (5968), 1008-1010. DOI: 10.1126/science.1182570 (acccessed 2025/04/17).

(3) Galloway, J. N.; Aber, J. D.; Erisman, J. W.; Seitzinger, S. P.; Howarth, R. W.; Cowling, E. B.; Cosby, B. J. The Nitrogen Cascade. *BioScience* **2003**, *53* (4), 341-356. DOI: 10.1641/0006-3568(2003)053[0341:TNC]2.0.CO;2 (acccessed 4/18/2025).

(4) Lelieveld, J.; Evans, J. S.; Fnais, M.; Giannadaki, D.; Pozzer, A. The contribution of outdoor air pollution sources to premature mortality on a global scale. *Nature* **2015**, *525* (7569), 367-371. DOI: 10.1038/nature15371.

(5) Gu, B.; Zhang, L.; Van Dingenen, R.; Vieno, M.; Van Grinsven, H. J. M.; Zhang, X.; Zhang, S.; Chen, Y.; Wang, S.; Ren, C.; et al. Abating ammonia is more cost-effective than nitrogen oxides for mitigating $PM_{2.5}$ air pollution. *Science* **2021**, *374* (6568), 758-762. DOI: 10.1126/science.abf8623 (acccessed 2025/04/17).

(6) Crippa, M.; Guizzardi, D.; Pagani, F.; Schiavina, M.; Melchiorri, M.; Pisoni, E.; Graziosi, F.; Muntean, M.; Maes, J.; Dijkstra, L.; et al. Insights into the spatial distribution of global, national, and subnational greenhouse gas emissions in the Emissions Database for Global Atmospheric Research (EDGAR v8.0). *Earth Syst. Sci. Data* **2024**, *16* (6), 2811-2830. DOI: 10.5194/essd-16-2811-2024.

(7) Kang, Y.; Liu, M.; Song, Y.; Huang, X.; Yao, H.; Cai, X.; Zhang, H.; Kang, L.; Liu, X.; Yan, X.; et al. High-resolution ammonia emissions inventories in China from 1980 to 2012. *Atmos. Chem. Phys.* **2016**, *16* (4), 2043-2058. DOI: 10.5194/acp-16-2043-2016.

(8) Li, B.; Chen, L.; Shen, W.; Jin, J.; Wang, T.; Wang, P.; Yang, Y.; Liao, H. Improved gridded ammonia emission inventory in China. *Atmos. Chem. Phys.* **2021**, *21* (20), 15883-15900. DOI: 10.5194/acp-21-15883-2021.

(9) Chen, S.; Cheng, M.; Guo, Z.; Xu, W.; Du, X.; Li, Y. Enhanced atmospheric ammonia ($NH_3$) pollution in China from 2008 to 2016: Evidence from a combination of observations and emissions. *Environ Pollut* **2020**, *263* (Pt B), 114421. DOI: 10.1016/j.envpol.2020.114421　From NLM Medline.

(10) Zhang, M.; Zhang, X.; Gao, C.; Zhao, H.; Zhang, S.; Xie, S.; Ran, L.; Xiu, A. Reactive nitrogen emissions from cropland and their dominant driving factors in China. *Science of The Total Environment* **2025**, *968*, 178919. DOI: https://doi.org/10.1016/j.scitotenv.2025.178919.

(11) Xu, P.; Zhang, Y.; Gong, W.; Hou, X.; Kroeze, C.; Gao, W.; Luan, S. An inventory of the emission of ammonia from agricultural fertilizer application in China for 2010 and its high-resolution spatial distribution. *Atmospheric Environment* **2015**, *115*, 141-148. DOI: https://doi.org/10.1016/j.atmosenv.2015.05.020.

(12) Liao, W.; Liu, M.; Huang, X.; Wang, T.; Xu, Z.; Shang, F.; Song, Y.; Cai, X.; Zhang, H.; Kang, L.; et al. Estimation for ammonia emissions at county level in China from 2013 to 2018. *Science China Earth Sciences* **2022**, *65* (6), 1116-1127. DOI: 10.1007/s11430-021-9897-3.

(13) Zheng, J. Y.; Yin, S. S.; Kang, D. W.; Che, W. W.; Zhong, L. J. Development and uncertainty analysis of a high-resolution $NH_3$ emissions inventory and its implications with precipitation over the Pearl River Delta region, China. *Atmos. Chem. Phys.* **2012**, *12* (15), 7041-7058. DOI: 10.5194/acp-12-7041-2012.

(14) Ma, S. High-resolution assessment of ammonia emissions in China: Inventories, driving forces and mitigation. *Atmospheric Environment* **2020**, *229*, 117458. DOI: https://doi.org/10.1016/j.atmosenv.2020.117458.

(15) Zhang, L.; Chen, Y.; Zhao, Y.; Henze, D. K.; Zhu, L.; Song, Y.; Paulot, F.; Liu, X.; Pan, Y.; Lin, Y.; et al. Agricultural ammonia emissions in China: reconciling bottom-up and top-down estimates. *Atmos. Chem. Phys.* **2018**, *18* (1), 339-355. DOI: 10.5194/acp-18-339-2018.

(16) Kong, L.; Tang, X.; Zhu, J.; Wang, Z.; Pan, Y.; Wu, H.; Wu, L.; Wu, Q.; He, Y.; Tian, S.; et al. Improved Inversion of Monthly Ammonia Emissions in China Based on the Chinese Ammonia Monitoring Network and Ensemble Kalman Filter. *Environmental Science & Technology* **2019**, *53* (21), 12529-12538. DOI: 10.1021/acs.est.9b02701.

(17) Luo, Z.; Zhang, Y.; Chen, W.; Van Damme, M.; Coheur, P.-F.; Clarisse, L. Estimating global ammonia ($NH_3$) emissions based on IASI observations from 2008 to 2018. *Atmos Chem Phys*. DOI: 10.5194/acp-2022-216.

(18) Jin, J.; Fang, L.; Li, B.; Liao, H.; Wang, Y.; Han, W.; Li, K.; Pang, M.; Wu, X.; Xiang Lin, H. 4DEnVar-based inversion system for ammonia emission estimation in China through assimilating IASI ammonia retrievals. *Environmental Research Letters* **2023**, *18* (3). DOI: 10.1088/1748-9326/acb835.

(19) Chen, Y.; Shen, H.; Kaiser, J.; Hu, Y.; Capps, S. L.; Zhao, S.; Hakami, A.; Shih, J.-S.; Pavur, G. K.; Turner, M. D.; et al. High-resolution hybrid inversion of IASI ammonia columns to constrain US ammonia emissions using the CMAQ adjoint model. *Atmospheric Chemistry and Physics* **2021**, *21* (3), 2067-2082. DOI: 10.5194/acp-21-2067-2021.

(20) Chen, J.; Du, X.; Liu, X.; Xu, W.; Krol, M. Estimation of Ammonia Emissions over China Using IASI Satellite-Derived Surface Observations. *Environmental Science & Technology* **2025**. DOI: 10.1021/acs.est.4c10878.

(21) Cao, H.; Henze, D. K.; Shephard, M. W.; Dammers, E.; Cady-Pereira, K.; Alvarado, M.; Lonsdale, C.; Luo, G.; Yu, F.; Zhu, L.; et al. Inverse modeling of $NH_3$ sources using CrIS remote sensing measurements. *Environmental Research Letters* **2020**, *15* (10), 104082. DOI: 10.1088/1748-9326/abb5cc.

(22) Ding, J.; van der A, R.; Eskes, H.; Dammers, E.; Shephard, M.; Wichink Kruit, R.; Guevara, M.; Tarrason, L. Ammonia emission estimates using CrIS satellite observations over Europe. *Atmospheric Chemistry and Physics* **2024**, *24* (18), 10583-10599. DOI: 10.5194/acp-24-10583-2024.

(23) Clarisse, L.; Shephard, M. W.; Dentener, F.; Hurtmans, D.; Cady-Pereira, K.; Karagulian, F.; Van Damme, M.; Clerbaux, C.; Coheur, P. F. Satellite monitoring of ammonia: A case study of the San Joaquin Valley. *Journal of Geophysical Research: Atmospheres* **2010**, *115* (D13). DOI: 10.1029/2009jd013291.

(24) Clarisse, L.; Franco, B.; Van Damme, M.; Di Gioacchino, T.; Hadji-Lazaro, J.; Whitburn, S.; Noppen, L.; Hurtmans, D.; Clerbaux, C.; Coheur, P. The IASI $NH_3$ version 4 product: averaging kernels and improved consistency. *Atmos. Meas. Tech.* **2023**, *16* (21), 5009-5028. DOI: 10.5194/amt-16-5009-2023.

(25) Van Damme, M.; Clarisse, L.; Heald, C. L.; Hurtmans, D.; Ngadi, Y.; Clerbaux, C.; Dolman, A. J.; Erisman, J. W.; Coheur, P. F. Global distributions, time series and error characterization of atmospheric ammonia ($NH_3$) from IASI satellite observations. *Atmos. Chem. Phys.* **2014**, *14* (6), 2905-2922. DOI: 10.5194/acp-14-2905-2014.

(26) Van Damme, M.; Whitburn, S.; Clarisse, L.; Clerbaux, C.; Hurtmans, D.; Coheur, P. F. Version 2 of the IASI $NH_3$ neural network retrieval algorithm: near-real-time and reanalysed datasets. *Atmos. Meas. Tech.* **2017**, *10* (12), 4905-4914. DOI: 10.5194/amt-10-4905-2017.

(27) Van Damme, M.; Clarisse, L.; Franco, B.; Sutton, M. A.; Erisman, J. W.; Wichink Kruit, R.; van Zanten, M.; Whitburn, S.; Hadji-Lazaro, J.; Hurtmans, D.; et al. Global, regional and national trends of atmospheric ammonia derived from a decadal (2008–2018) satellite record. *Environmental Research Letters* **2021**, *16* (5). DOI: 10.1088/1748-9326/abd5e0.

(28) Whitburn, S.; Van Damme, M.; Clarisse, L.; Bauduin, S.; Heald, C. L.; Hadji-Lazaro, J.; Hurtmans, D.; Zondlo, M. A.; Clerbaux, C.; Coheur, P. F. A flexible and robust neural network IASI-$NH_3$ retrieval algorithm. *Journal of Geophysical Research: Atmospheres* **2016**, *121* (11), 6581-6599. DOI: 10.1002/2016jd024828.

(29) Byun, D.; Schere, K. L. Review of the Governing Equations, Computational Algorithms, and Other Components of the Models-3 Community Multiscale Air Quality (CMAQ) Modeling System. *Applied Mechanics Reviews* **2006**, *59* (2), 51-77. DOI: 10.1115/1.2128636 (acccessed 4/19/2025).

(30) US EPA Office of Research and Development. CMAQv5.0 (5.0). *Zenodo* **2012**. DOI: https://doi.org/10.5281/zenodo.1079888.

(31) Zhao, S.; Russell, M. G.; Hakami, A.; Capps, S. L.; Turner, M. D.; Henze, D. K.; Percell, P. B.; Resler, J.; Shen, H.; Russell, A. G.; et al. A multiphase CMAQ version 5.0 adjoint. *Geosci. Model Dev.* **2020**, *13* (7), 2925-2944. DOI: 10.5194/gmd-13-2925-2020.

(32) Fountoukis, C.; Nenes, A. ISORROPIA II: a computationally efficient thermodynamic equilibrium model for $K^+$–$Ca^{2+}$–$Mg^{2+}$–$NH_4^+$–$Na^+$–$SO_4^{2-}$–$NO_3^-$–$Cl^-$–$H_2O$ aerosols. *Atmos. Chem. Phys.* **2007**, *7* (17), 4639-4659. DOI: 10.5194/acp-7-4639-2007.

(33) Yarwood, G.; Rao, S.; Yocke, M.; Whitten, G. Z. Updates to the Carbon Bond Mechanism: CB05. **2005**.

(34) Zheng, L.; Adalibieke, W.; Zhou, F.; He, P.; Chen, Y.; Guo, P.; He, J.; Zhang, Y.; Xu, P.; Wang, C.; et al. Health burden from food systems is highly unequal across income groups. *Nature Food* **2024**, *5* (3), 251-261. DOI: 10.1038/s43016-024-00946-7.

(35) National Centers for Environmental Prediction NCEP Products Inventory, Global Products, Global Forecast System (GFS) Model. https://www.nco.ncep.noaa.gov/pmb/products/gfs/#GFS.

(36) Clarisse, L.; Clerbaux, C.; Dentener, F.; Hurtmans, D.; Coheur, P.-F. Global ammonia distribution derived from infrared satellite observations. *Nature Geoscience* **2009**, *2* (7), 479-483. DOI: 10.1038/ngeo551.

(37) Xu, P.; Koloutsou-Vakakis, S.; Rood, M. J.; Luan, S. Projections of $NH_3$ emissions from manure generated by livestock production in China to 2030 under six mitigation scenarios. *Science of The Total Environment* **2017**, *607-608*, 78-86. DOI: https://doi.org/10.1016/j.scitotenv.2017.06.258.

(38) AiMa Forecasts AiMa Air Quality Forecasting System. http://www.aimayubao.com.

(39) Shen, H.; Sun, Z.; Chen, Y.; Russell, A. G.; Hu, Y.; Odman, M. T.; Qian, Y.; Archibald, A. T.; Tao, S. Novel Method for Ozone Isopleth Construction and Diagnosis for the Ozone Control Strategy of Chinese Cities. *Environmental Science & Technology* **2021**, *55* (23), 15625-15636. DOI: 10.1021/acs.est.1c01567.

(40) Zhang, Q.; Streets, D. G.; Carmichael, G. R.; He, K. B.; Huo, H.; Kannari, A.; Klimont, Z.; Park, I. S.; Reddy, S.; Fu, J. S.; et al. Asian emissions in 2006 for the NASA INTEX-B mission. *Atmos. Chem. Phys.* **2009**, *9* (14), 5131-5153. DOI: 10.5194/acp-9-5131-2009.

(41) Paulot, F.; Jacob, D. J.; Pinder, R. W.; Bash, J. O.; Travis, K.; Henze, D. K. Ammonia emissions in the United States, European Union, and China derived by high-resolution inversion of ammonium wet deposition data: Interpretation with a new agricultural emissions inventory (MASAGE_$NH_3$). *Journal of Geophysical Research: Atmospheres* **2014**, *119* (7), 4343-4364. DOI: https://doi.org/10.1002/2013JD021130.

(42) Hansen, P. C. The L-Curve and its Use in the Numerical Treatment of Inverse Problems. In *Computational Inverse Problems in Electrocardiology*, Advances in Computational Bioengineering, Vol. 5; WIT Press, 2001; pp 119-142.

(43) Pan, Y.; Tian, S.; Zhao, Y.; Zhang, L.; Zhu, X.; Gao, J.; Huang, W.; Zhou, Y.; Song, Y.; Zhang, Q.; et al. Identifying Ammonia Hotspots in China Using a National Observation Network. *Environ Sci Technol* **2018**, *52* (7), 3926-3934. DOI: 10.1021/acs.est.7b05235  From NLM Medline.

(44) Fu, H.; Luo, Z.; Hu, S. A temporal-spatial analysis and future trends of ammonia emissions in China. *Science of The Total Environment* **2020**, *731*, 138897. DOI: https://doi.org/10.1016/j.scitotenv.2020.138897.

(45) Liu, P.; Ding, J.; Liu, L.; Xu, W.; Liu, X. Estimation of surface ammonia concentrations and emissions in China from the polar-orbiting Infrared Atmospheric Sounding Interferometer and the FY-4A Geostationary

Interferometric Infrared Sounder. *Atmospheric Chemistry and Physics* **2022**, *22* (13), 9099-9110. DOI: 10.5194/acp-22-9099-2022.

(46) Chen P L, X. X. X., Wang Q G. High-resolution characteristics of $NH_3$ emission from 2010 to 2020 in China based on satellite observation [J]. *China Environmental Science* **2023**, *43* (6), 2673-2682.

(47) Geng, G.; Liu, Y.; Liu, Y.; Liu, S.; Cheng, J.; Yan, L.; Wu, N.; Hu, H.; Tong, D.; Zheng, B.; et al. Efficacy of China's clean air actions to tackle $PM_{2.5}$ pollution between 2013 and 2020. *Nature Geoscience* **2024**, *17* (10), 987-994. DOI: 10.1038/s41561-024-01540-z.

(48) Crippa, M.; Guizzardi, D.; Butler, T.; Keating, T.; Wu, R.; Kaminski, J.; Kuenen, J.; Kurokawa, J.; Chatani, S.; Morikawa, T.; et al. The HTAP_v3 emission mosaic: merging regional and global monthly emissions (2000–2018) to support air quality modelling and policies. *Earth Syst. Sci. Data* **2023**, *15* (6), 2667-2694. DOI: 10.5194/essd-15-2667-2023.

(49) Huang, X.; Song, Y.; Li, M.; Li, J.; Huo, Q.; Cai, X.; Zhu, T.; Hu, M.; Zhang, H. A high-resolution ammonia emission inventory in China. *Global Biogeochemical Cycles* **2012**, *26* (1). DOI: 10.1029/2011gb004161.

(50) Hoesly, R., Smith, S. J., Prime, N., Ahsan, H., Suchyta, H., O'Rourke, P., Crippa, M., Klimont, Z., Guizzardi, D., Behrendt, J., Feng, L., Harkins, C., McDonald, B., Mott, A., McDuffie, A., Nicholson, M., & Wang, S. CEDS v_2024_07_08 Release Emission Data. v_2024_07_08 ed.; Zenodo, 2024.

(51) Evangeliou, N.; Balkanski, Y.; Eckhardt, S.; Cozic, A.; Van Damme, M.; Coheur, P. F.; Clarisse, L.; Shephard, M. W.; Cady-Pereira, K. E.; Hauglustaine, D. 10-year satellite-constrained fluxes of ammonia improve performance of chemistry transport models. *Atmos. Chem. Phys.* **2021**, *21* (6), 4431-4451. DOI: 10.5194/acp-21-4431-2021.

(52) Zhang, X.; Wu, Y.; Liu, X.; Reis, S.; Jin, J.; Dragosits, U.; Van Damme, M.; Clarisse, L.; Whitburn, S.; Coheur, P. F.; et al. Ammonia Emissions May Be Substantially Underestimated in China. *Environ Sci Technol* **2017**, *51* (21), 12089-12096. DOI: 10.1021/acs.est.7b02171.

(53) Wang, H.; Zhang, D.; Zhang, Y.; Zhai, L.; Yin, B.; Zhou, F.; Geng, Y.; Pan, J.; Luo, J.; Gu, B.; et al. Ammonia emissions from paddy fields are underestimated in China. *Environ Pollut* **2018**, *235*, 482-488. DOI: 10.1016/j.envpol.2017.12.103.

(54) Zhang, X.; Gu, B.; van Grinsven, H.; Lam, S. K.; Liang, X.; Bai, M.; Chen, D. Societal benefits of halving agricultural ammonia emissions in China far exceed the abatement costs. *Nature Communications* **2020**, *11* (1), 4357. DOI: 10.1038/s41467-020-18196-z.

(55) Zheng, B.; Tong, D.; Li, M.; Liu, F.; Hong, C.; Geng, G.; Li, H.; Li, X.; Peng, L.; Qi, J.; et al. Trends in China's anthropogenic emissions since 2010 as the consequence of clean air actions. *Atmos. Chem. Phys.* **2018**, *18* (19), 14095-14111. DOI: 10.5194/acp-18-14095-2018.

(56) Shephard, M. W.; Cady-Pereira, K. E. Cross-track Infrared Sounder (CrIS) satellite observations of tropospheric ammonia. *Atmos. Meas. Tech.* **2015**, *8* (3), 1323-1336. DOI: 10.5194/amt-8-1323-2015.

(57) Shephard, M. W.; Cady-Pereira, K. E.; Luo, M.; Henze, D. K.; Pinder, R. W.; Walker, J. T.; Rinsland, C. P.; Bash, J. O.; Zhu, L.; Payne, V. H.; et al. TES ammonia retrieval strategy and global observations of the spatial and seasonal variability of ammonia. *Atmospheric Chemistry and Physics* **2011**, *11* (20), 10743-10763. DOI: 10.5194/acp-11-10743-2011.

(58) Zhang, Y.; Tang, A.; Wang, D.; Wang, Q.; Benedict, K.; Zhang, L.; Liu, D.; Li, Y.; Collett Jr, J. L.; Sun, Y.; et al. The vertical variability of ammonia in urban Beijing, China. *Atmos. Chem. Phys.* **2018**, *18* (22), 16385-16398. DOI: 10.5194/acp-18-16385-2018.

(59) Pu, W.; Guo, H.; Ma, Z.; Qiu, Y.; Tang, Y.; Liu, Q.; Wang, F.; Sheng, J. Aircraft measurements reveal vertical distribution of atmospheric ammonia over the North China Plain in early autumn. *Environmental Chemistry Letters* **2020**, *18* (6), 2149-2156. DOI: 10.1007/s10311-020-01051-4.

(60) Zhang, Y.; Tang, A.; Wang, D.; Wang, Q.; Benedict, K.; Zhang, L.; Liu, D.; Li, Y.; Collett Jr, J. L.; Sun, Y.; et al. The vertical variability of ammonia in urban Beijing, China. *Atmospheric Chemistry and Physics* **2018**, *18* (22), 16385-16398. DOI: 10.5194/acp-18-16385-2018.

(61) Wang, D.; Huo, J.; Duan, Y.; Zhang, K.; Ding, A.; Fu, Q.; Luo, J.; Fei, D.; Xiu, G.; Huang, K. Vertical distribution and transport of air pollutants during a regional haze event in eastern China: A tethered mega-balloon observation study. *Atmospheric Environment* **2021**, *246*. DOI: 10.1016/j.atmosenv.2020.118039.

(62) Höpfner, M.; Ungermann, J.; Borrmann, S.; Wagner, R.; Spang, R.; Riese, M.; Stiller, G.; Appel, O.; Batenburg, A. M.; Bucci, S.; et al. Ammonium nitrate particles formed in upper troposphere from ground ammonia sources during Asian monsoons. *Nature Geoscience* **2019**, *12* (8), 608-612. DOI: 10.1038/s41561-019-0385-8.

(63) Pawar, P. V.; Ghude, S. D.; Jena, C.; Móring, A.; Sutton, M. A.; Kulkarni, S.; Lal, D. M.; Surendran, D.; Van Damme, M.; Clarisse, L.; et al. Analysis of atmospheric ammonia over South and East Asia based on the MOZART-4 model and its comparison with satellite and surface observations. *Atmos. Chem. Phys.* **2021**, *21* (8), 6389-6409. DOI: 10.5194/acp-21-6389-2021.

(64) Clarisse, L.; Franco, B.; Van Damme, M.; Di Gioacchino, T.; Hadji-Lazaro, J.; Whitburn, S.; Noppen, L.; Hurtmans, D.; Clerbaux, C.; Coheur, P. The IASI $NH_3$ version 4 product: averaging kernels and improved consistency. DOI: 10.5194/amt-2023-48.

(65) *Regional Classification in China*. Resource and Environment Science and Data Center, 2014. http://www.resdc.cn/data.aspx?DATAID=276 (accessed 2025 April 21).

(66) Liu, X.; Zhang, W.; Hu, Y.; Hu, E.; Xie, X.; Wang, L.; Cheng, H. Arsenic pollution of agricultural soils by concentrated animal feeding operations (CAFOs). *Chemosphere* **2015**, *119*, 273-281. DOI: https://doi.org/10.1016/j.chemosphere.2014.06.067.

(67) Carter, C. A. China's Agriculture Achievements and Challenges. *ARE Update* **2011**, *14*, 5-7.

(68) Sha WEI, J. F., Yanfeng TIAN, Hongmin DONG. Low-carbon development policies and achievements in the context of the livestock sector in China. *Front. Agr. Sci. Eng.* **2024**, *11* (3), 367-380. DOI: 10.15302/j-fase-2024553.

(69) Sauter, C.; Grether, J.-M.; Mathys, N. A. Geographical spread of global emissions: Within-country inequalities are large and increasing. *Energy Policy* **2016**, *89*, 138-149. DOI: https://doi.org/10.1016/j.enpol.2015.11.024.

(70) Wyer, K. E.; Kelleghan, D. B.; Blanes-Vidal, V.; Schauberger, G.; Curran, T. P. Ammonia emissions from agriculture and their contribution to fine particulate matter: A review of implications for human health. *Journal of Environmental Management* **2022**, *323*, 116285. DOI: https://doi.org/10.1016/j.jenvman.2022.116285.

(71) Duan, J.; Liu, H.; Zhang, X.; Ren, C.; Wang, C.; Cheng, L.; Xu, J.; Gu, B. Agricultural management practices in China enhance nitrogen sustainability and benefit human health. *Nature Food* **2024**, *5* (5), 378-389. DOI: 10.1038/s43016-024-00953-8.

(72) Dammers, E.; Palm, M.; Van Damme, M.; Vigouroux, C.; Smale, D.; Conway, S.; Toon, G. C.; Jones, N.; Nussbaumer, E.; Warneke, T.; et al. An evaluation of IASI-$NH_3$ with ground-based Fourier transform infrared spectroscopy measurements. *Atmos. Chem. Phys.* **2016**, *16* (16), 10351-10368. DOI: 10.5194/acp-16-10351-2016.

(73) Lyu, X.; Zeng, Y.; Tian, S.; Sun, J.; Zhang, G.; Huang, W.; Gu, M.; Xu, W.; Liu, X.; Dong, H.; et al. Atmospheric reactive nitrogen in typical croplands and intensive pig and poultry farms in the North China Plain. *Chinese Journal of Eco-Agriculture* **2020**, *28* (7), 1043-1050. DOI: 10.13930/j.cnki.cjea.190879.

*Supporting Information*

# Vertical Profile Corrected Satellite NH₃ Retrievals Enable Accurate Agricultural Emission Characterization in China

Qiming Liu[1,2], Yilin Chen[3,*], Peng Xu[4], Huizhong Shen[1,2], Zelin Mai[1,2], Ruixin Zhang[1,2], Peng Guo[1,2], Zhiyu Zheng[1,2], Tiancheng Luan[1,2], Shu Tao[1,2,5,6]

[1]Shenzhen Key Laboratory of Precision Measurement and Early Warning Technology for Urban Environmental Health Risks, School of Environmental Science and Engineering, Southern University of Science and Technology, Shenzhen 518055, China

[2]Guangdong Provincial Observation and Research Station for Coastal Atmosphere and Climate of the Greater Bay Area, School of Environmental Science and Engineering, Southern University of Science and Technology, Shenzhen 518055, China

[3]School of Urban Planning and Design, Peking University, Shenzhen Graduate School, Shenzhen 518055, China

[4]Institute of Surface–Earth System Science, School of Earth System Science, Tianjin University, Tianjin 300072, China

[5]College of Urban and Environmental Sciences, Peking University, Beijing 100871, China

[6]Institute of Carbon Neutrality, Peking University, Beijing 100871, China

*Corresponding author, e-mail: ylchen2023@pku.edu.cn

**Contents**

Text S1. Pre- and Post-filtering in the Reprocessing of IASI NH₃

Text S2. Iterative Mass Balance (IMB) Method

Text S3. Sensitivity Test: Contribution of Regional Transport to Elevated NH₃ Profiles

Figure S1. Spatial distribution of monitoring stations and regional classification used in this study.

Figure S2. Evaluation of WRF meteorological variables against hourly surface observations for January, April, May, June, July, and October.

Figure S3. L-curves from the first iteration for January, April, May, June, July, and October used to determine the regularization factor ($\gamma$).

Figure S4. Seasonal variation in simulated $NH_3$ vertical profiles and annual averaging kernel characteristics over China.

Figure S5. Impact of emission scenarios on $NH_3$ vertical profiles over Northwest and Southwest China in April.

Figure S6. $NH_3$ total column density patterns from CMAQ simulations and model evaluation for May and June.

Figure S7. Comparison of surface $NH_3$ concentrations simulated by CMAQ using a priori and posterior emission inventories for May and June.

Figure S8. Comparison of prior and posterior CMAQ-simulated surface $NH_3$ concentrations with surface observations in high-emission regions.

Figure S9. Comparison of surface $NH_3$ observations with CMAQ-simulated surface $NH_3$ concentrations based on the baseline IASI inversion inventory.

Table S1. Comparison of CMAQ vertical profiles and IASI $NH_3$ baseline vertical profiles with surface observations in two regions.

**Text S1.** Pre- and Post-filtering in the Reprocessing of IASI NH$_3$

Prior to averaging kernel (AVK) calculations, a pre-filter provided by the satellite product was applied to eliminate retrievals ($X^a$) affected by erroneous Level 1 signals or excessive cloud cover. Additionally, our recalculate post-filter flagged retrievals that exhibited either: (1) limited or no sensitivity to the measured quantity, or (2) HRIs that were either too noisy or incompatible with the assumed vertical profile.

$$\frac{X^m}{HRI} > 1.5 \times 10^{16} \, molec \cdot cm^{-2} \quad \text{(condition 1)}$$

$$X^a < 0 \quad \text{(condition 2)}$$

$$|HRI| > 1.5 \quad \text{(condition 3)}$$

After computing the adjusted retrievals ($X^m$) via AVK correction, a post-filtering step was essential to reassess $X^m$, removing data points that either meet condition 1 or simultaneously satisfy conditions 2 and 3. The resulting filtered $X^m$ constitutes the final reprocessed IASI NH$_3$ dataset.

**Text S2.** Iterative Mass Balance (IMB) Method

The IMB method was first executed at coarse resolution, performing monthly-scale adjustments based on the ratio of IASI $NH_3$ to CMAQ-simulated monthly average ammonia column densities. This ratio was applied iteratively to scale emissions and obtain optimal IMB estimates. Given the atmospheric lifetime of $NH_3$—ranging from several to over a dozen hours – the procedure was conducted on a coarse 216 km by 216 km grid to efficiently minimize the coupling between local emissions and regional transport.

To ensure full spatial coverage during downscaling from the full simulation domain (124 by 184 grid cells at 36 km by 36 km resolution) to the coarser 216 km by 216 km resolution, we employed a moving window approach. This method anchored iterative updates at each of the domain's four corners, preventing the omission of any grid cells and ensuring comprehensive domain-wide inversion.

In each iteration, emissions in a grid cell were scaled proportionally using the ratio of observed to simulated $NH_3$ column concentrations, following:

$$E_t = E_a \times \frac{\Omega_o}{\Omega_a}$$

where $E_t$ and $E_a$ represent the grid-level emissions before and after adjustment, and $\Omega_a$ and $\Omega_o$ are the CMAQ and IASI monthly average $NH_3$ column concentrations, respectively. Emissions updated in each iteration served as inputs for CMAQ to simulate new $NH_3$ columns, which were then used to calculate the next adjustment. The final converged emissions were used as the prior input for the subsequent 4D-Var inversion.

**Text S3.** Sensitivity Test: Contribution of Regional Transport to Elevated $NH_3$ Profiles

To assess the contribution of regional transport to elevated $NH_3$ concentrations above the planetary boundary layer (PBL), we performed sensitivity simulations in April for two key regions—Northwest (Xinjiang) and Southwest (Tibet) China. In these hypothetical scenarios (NW_T and SW_T), $NH_3$ emissions from India and lateral boundary conditions (BCON) were set to zero, isolating the effect of cross-boundary transport. These were compared against realistic emission scenarios (NW and SW) with all emissions retained. Results reveal substantial reductions in above-PBL $NH_3$ concentrations—up to 24 percentage points—highlighting the significant influence of long-range transport on vertical $NH_3$ profiles in western China (Figure S5). This finding underscores the importance of incorporating regional transport in satellite retrieval interpretation and emission inversion.

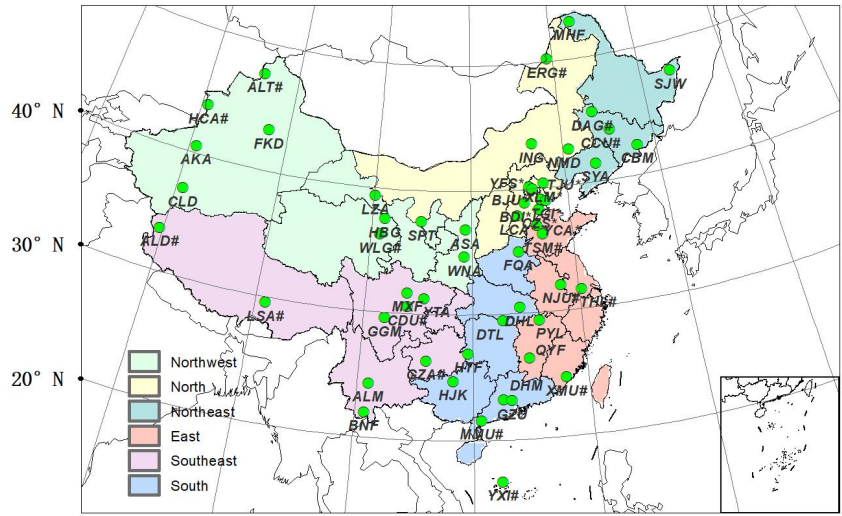

**Figure S1.** Spatial distribution of monitoring stations and regional classification used in this study. The locations of 53 monitoring stations, obtained from the AMoN-China network, are shown as green dots, with station codes labeled beneath each point. The study area is divided into six regions—East, South, Northwest, North, Northeast, and Southwest—based on the regional delineation provided by the Resource and Environment Science Data Center.[1]

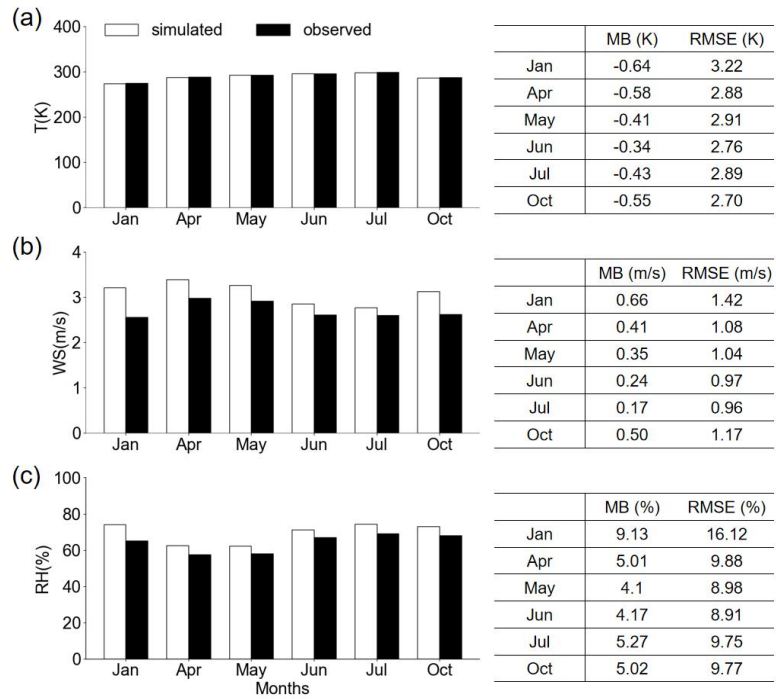

**Figure S2.** Evaluation of WRF meteorological variables against hourly surface observations for January, April, May, June, July, and October. Panels show comparisons between model simulations (white bars) and observed data (black bars) for (a) temperature (T, K), (b) wind speed (WS, m/s), and (c) relative humidity (RH, %). Corresponding mean bias (MB) and root mean square error (RMSE) values are presented alongside each panel to support quantitative assessment of model performance.

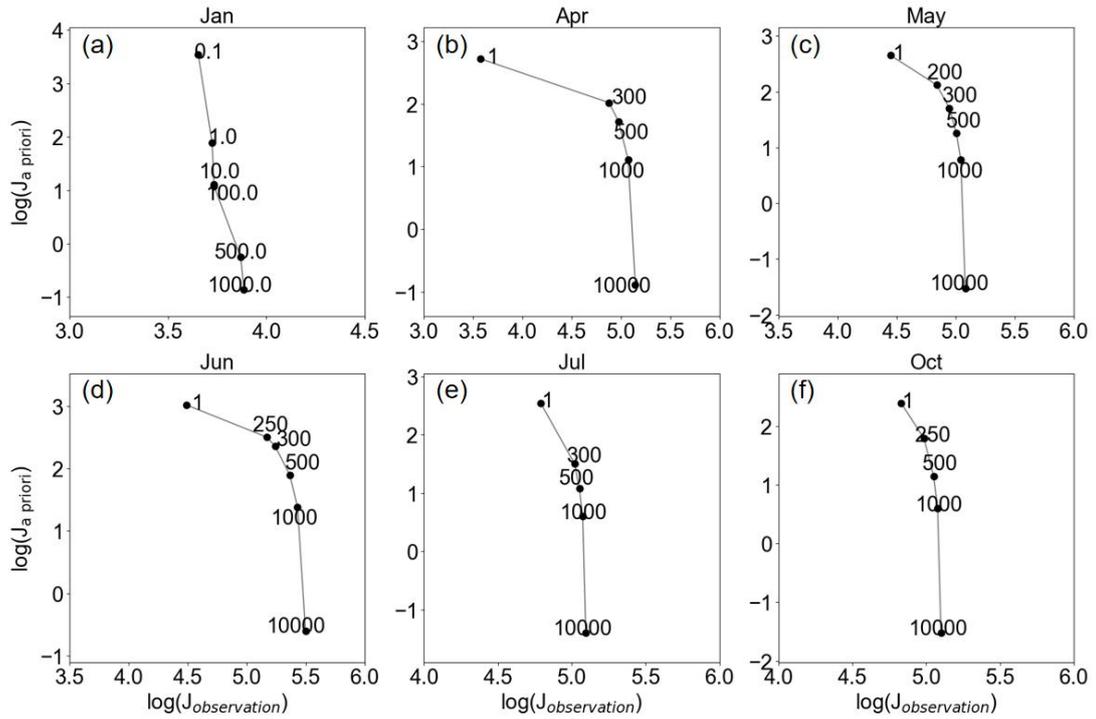

**Figure S3.** L-curves from the first iteration for January, April, May, June, July, and October used to determine the regularization factor ($\gamma$). Each curve depicts the trade-off between the error-weighted squared deviation of emission scaling factors from their a priori values ($J_{a\,priori}$) and the error-weighted squared mismatch between IASI-NH$_3$ observations and simulated column densities ($J_{observation}$) across a range of $\gamma$ values. In the first iteration, $\gamma$ was initialized as follows: 1 for January, 300 for April, 200 for May, 300 for June, 300 for July, and 250 for October. In subsequent iterations, $\gamma$ was adaptively updated based on the shape of the corresponding L-curve.

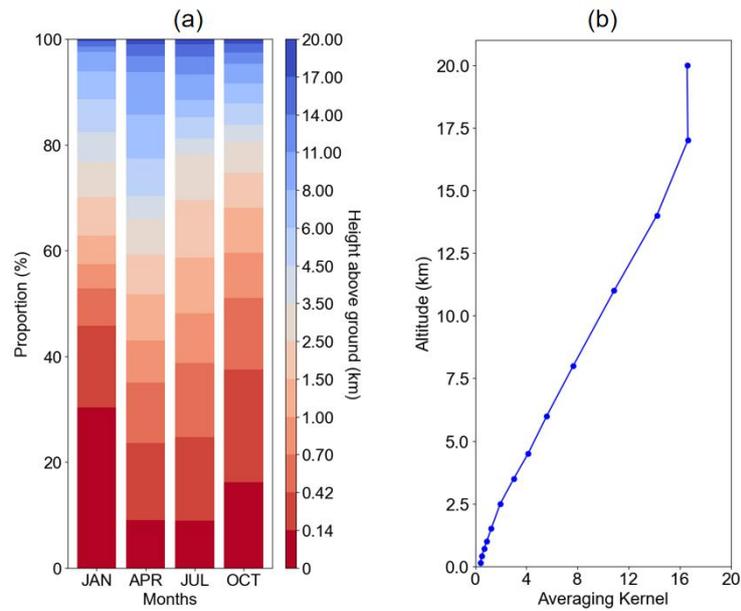

**Figure S4.** Seasonal variation in simulated NH$_3$ vertical profiles and annual averaging kernel characteristics over China. (a) Monthly mean vertical profiles of ammonia (NH3) simulated by CMAQ for January, April, July, and October, depicting seasonal variations across the study region. Colors represent the proportion of NH3 concentration at different altitudes above ground. (b) Annual mean averaging kernel values as a function of altitude, indicating the vertical sensitivity of NH$_3$ retrievals over China.[2]

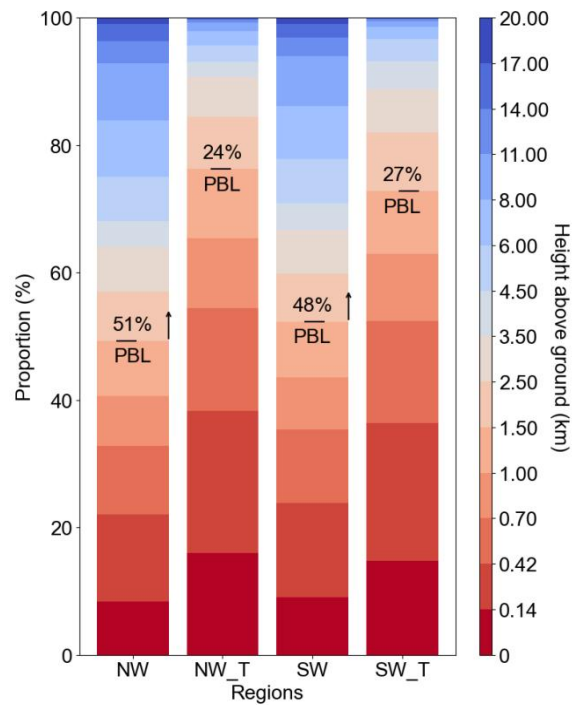

**Figure S5.** Impact of emission scenarios on NH$_3$ vertical profiles over Northwest and Southwest China in April. Vertical profiles of ammonia (NH$_3$) simulated by CMAQ are shown for two regions under two emission scenarios: realistic (NW and SW) and hypothetical (NW_T and SW_T), where NH$_3$ emissions from India and lateral boundary inputs were removed in the latter to isolate long-range transport effects. Percentages above each bar indicate the fraction of NH$_3$ concentration residing above the planetary boundary layer (PBL) relative to the total column.

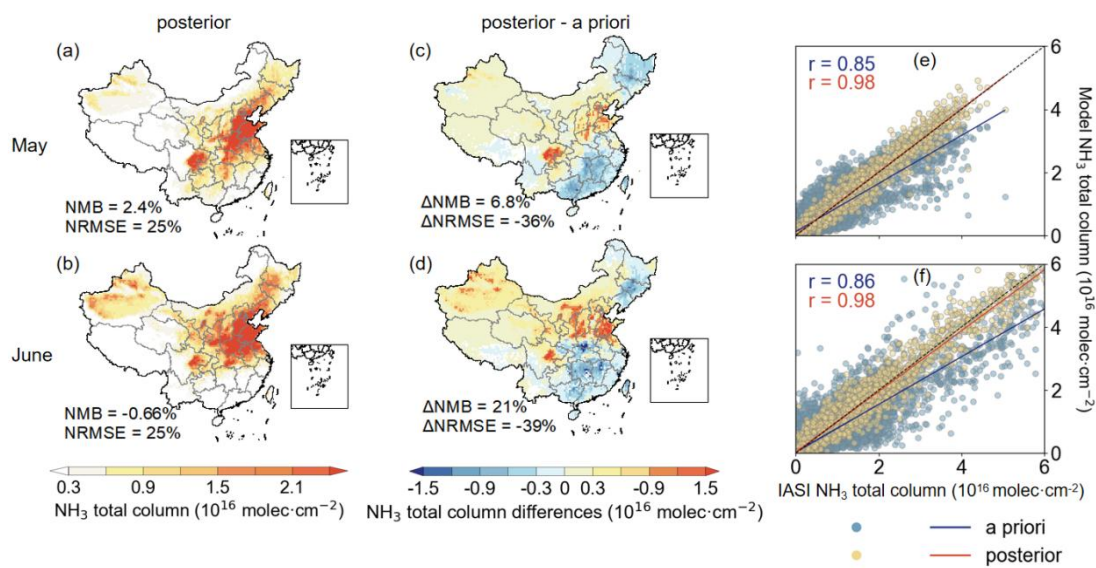

**Figure S6.** $NH_3$ total column density patterns from CMAQ simulations and model evaluation for May and June. (a–b) CMAQ-simulated $NH_3$ column densities using the posterior emission inventory for May and June across China. Normalized mean bias (NMB) and normalized root-mean-square error (NRMSE) relative to reprocessed IASI observations shown. (c–d) Spatial differences between CMAQ simulations using posterior and a priori emission inventories for the same months, with corresponding changes in performance metrics (ΔNMB and ΔNRMSE). (e–f) Scatter plots comparing reprocessed IASI $NH_3$ columns with CMAQ simulations using a priori (blue) and posterior (red) emission inventories, with Pearson correlation coefficients (r) for each simulation.

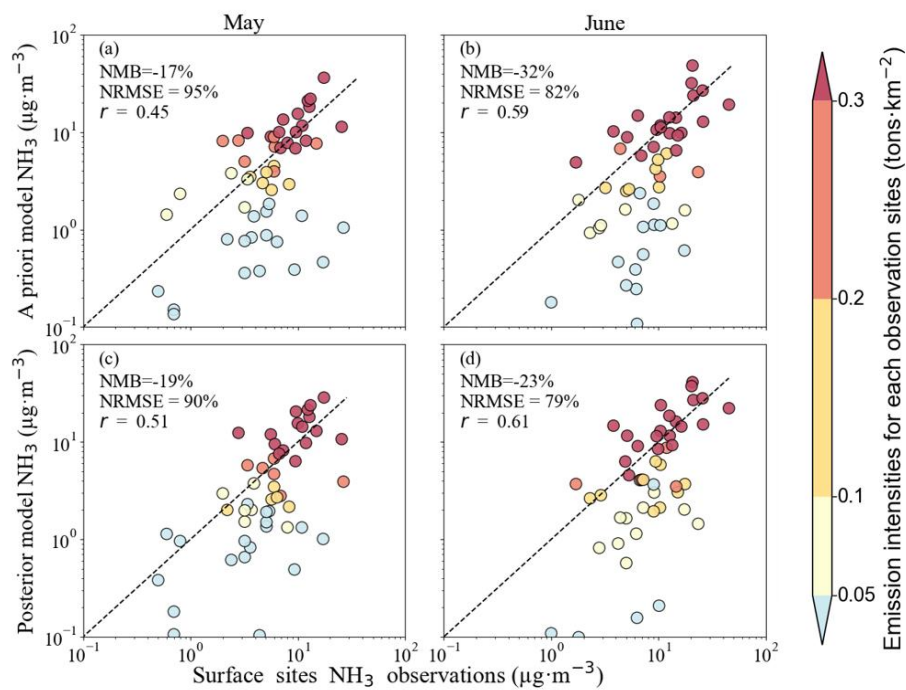

**Figure S7.** Comparison of surface NH₃ concentrations simulated by CMAQ using a priori and posterior emission inventories for May and June. (a–b) Surface NH₃ concentrations simulated using the prior emission inventory. (c–d) Same as (a–b), but using the posterior emission inventory. All simulations are evaluated against independent surface observations from the AMoN-China network. The color intensity of each point represents the prior emission magnitude at the corresponding observation site for the given month. Normalized mean bias (NMB), normalized root-mean-square error (NRMSE), and Pearson correlation coefficient ($r$) across all sites are shown in the upper left corner of each panel.

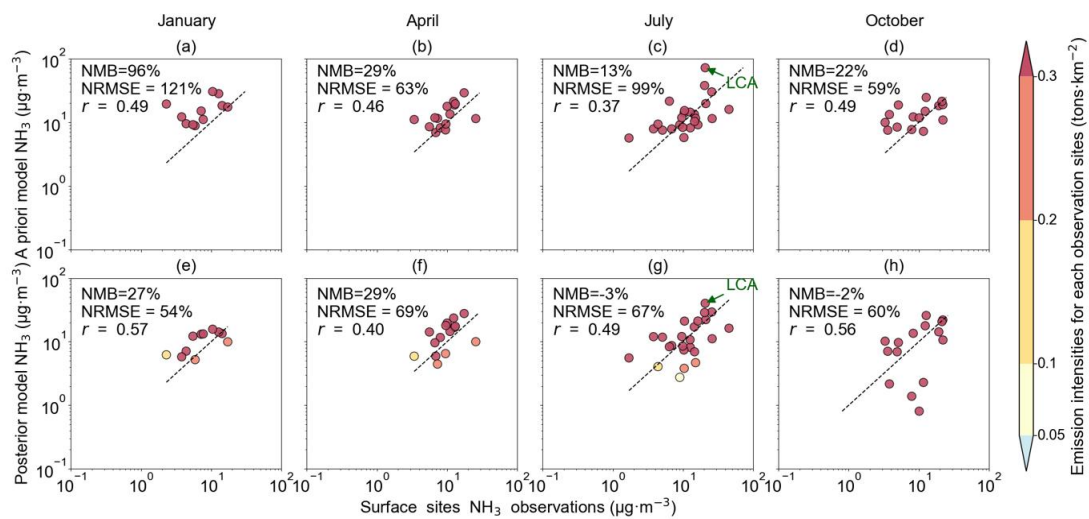

**Figure S8.** Comparison of prior and posterior CMAQ-simulated surface NH$_3$ concentrations with surface observations in high-emission regions. High-emission regions are defined as grid cells where monthly a priori NH$_3$ emission intensities exceed 0.3 tons·km$^{-2}$. Panels (a–d) show simulations using the prior emission inventory, while panels (e–h) show simulations using the posterior emission inventory. Model performance is evaluated against surface observations to assess accuracy in areas characterized by intensive agricultural activity. Normalized mean bias (NMB), normalized root-mean-square error (NRMSE), and Pearson correlation coefficient (r) are provided in each panel.

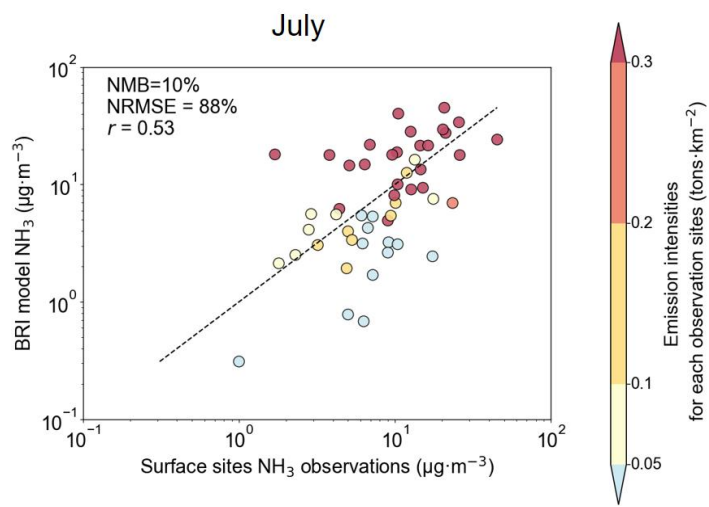

**Figure S9.** Comparison of surface $NH_3$ observations with CMAQ-simulated surface $NH_3$ concentrations based on the baseline IASI inversion inventory. The baseline inversion inventory is derived from prior IASI retrievals assuming a fixed vertical profile during the inversion. Corresponding surface $NH_3$ concentrations simulated by CMAQ are evaluated against ground-based observations. Normalized mean bias (NMB), normalized root-mean-square error (NRMSE), and Pearson correlation coefficient (r) are reported in the upper left corner.

| regions | Hight (m) | Observation | CMAQ | IASI-NH$_3$ baseline vertical profile |
|---|---|---|---|---|
| Beijing[3] | 2-320 | 1.1:1 | 2.3:1 | 3.3:1 |
| Baoding[4] | 500-3500 | 2.5:1 | 9.6:1 | 13:1 |
| | 500-1500 | 1.4:1 | 2.4:1 | 4.8:1 |
| | 1500-3500 | 1.7:1 | 4.0:1 | 25:1 |

Assuming the PBLH is 1500 meters.

**Table S1.** Comparison of CMAQ vertical profiles and IASI NH$_3$ baseline vertical profiles with surface observations in two regions. Ratios represent NH$_3$ concentrations at specified altitude ranges relative to reference surface values. Heights are given in meters, with planetary boundary layer height (PBLH) assumed at 1500 m.

# References

(1) *Regional Classification in China*. Resource and Environment Science and Data Center, 2014. http://www.resdc.cn/data.aspx?DATAID=276 (accessed 2025 April 21).

(2) Clarisse, L.; Franco, B.; Van Damme, M.; Di Gioacchino, T.; Hadji-Lazaro, J.; Whitburn, S.; Noppen, L.; Hurtmans, D.; Clerbaux, C.; Coheur, P. The IASI $NH_3$ version 4 product: averaging kernels and improved consistency. *Atmos. Meas. Tech.* **2023**, *16* (21), 5009-5028. DOI: 10.5194/amt-16-5009-2023.

(3) Zhang, Y.; Tang, A.; Wang, D.; Wang, Q.; Benedict, K.; Zhang, L.; Liu, D.; Li, Y.; Collett Jr, J. L.; Sun, Y.; et al. The vertical variability of ammonia in urban Beijing, China. *Atmos. Chem. Phys.* **2018**, *18* (22), 16385-16398. DOI: 10.5194/acp-18-16385-2018.

(4) Pu, W.; Guo, H.; Ma, Z.; Qiu, Y.; Tang, Y.; Liu, Q.; Wang, F.; Sheng, J. Aircraft measurements reveal vertical distribution of atmospheric ammonia over the North China Plain in early autumn. *Environmental Chemistry Letters* **2020**, *18* (6), 2149-2156. DOI: 10.1007/s10311-020-01051-4.